\documentclass[showpacs,tightenlines,nofootinbib,twocolumn,aps,prd,amsmath,amssymb,floatfix]{revtex4}
\usepackage{graphics,bm}
\usepackage{epsfig}
\usepackage{graphicx}

\newcommand{\beq}{\begin{equation}}
\newcommand{\eeq}{\end{equation}}
\newcommand{\bqa}{\begin{eqnarray}}
\newcommand{\eqa}{\end{eqnarray}}

\begin{document}

\pagestyle{plain}
\font\tenrm=cmr10
\def\sumint{\hbox{$\sum$}\!\!\!\!\!\!\int}
\def\square{\vcenter{\vbox{\hrule height.4pt
          \hbox{\vrule width.4pt height4pt
          \kern4pt\vrule width.3pt}\hrule height.4pt}}}
\def\boxx{\square}

\title{$1/N$-Expansion and the Dilute Bose Gas Beyond Mean-field Theory}
\author{Jens O. Andersen}
\email{jensoa@ntnu.no}

\affiliation{Department of Physics, Norwegian Institute of Science and 
Technology, N-7491 Trondheim, Norway}

\date{\today}

\begin{abstract}
We consider the nonrelativistic interacting Bose gas at zero and finite
temperature. Using the $1/N$-expansion, we derive expressions for the
free energy density and the number density to next-to-leading order in $1/N$.
Outside the critical region and at weak coupling, our
calculations reduce to the well-known mean-field results for the dilute
Bose gas. We also rederive the nonperturbative
critical density for Bose condensation which 
was first calculated by Baym, Blaizot, Holzmann, Lal\"oe, and Vautherin.
\end{abstract}


\maketitle

\small

\section{Introduction}
The dilute Bose gas has a very long history dating back to the classic
paper by Bogoliubov in 1947~\cite{bogo1}. At zero temperature, the
quantum loop expansion is essentially an expansion in powers of the
gas parameter $\sqrt{\rho a^3}$, where $\rho$ is the number density and $a$
is the $s$-wave scattering length. The leading quantum corrections to
the mean-field results for
the energy density and the speed of sound were calculated many years ago
by Lee and Yang~\cite{leeyang1}.

Finite-temperature corrections to the pressure of a dilute Bose gas
were first calculated by Lee and Yang~\cite{leeyang2}.
By performing an expansion of the pressure about zero temperature, they
showed that the leading correction term goes as $T^4$. This reflects the
fact that
the thermodynamics at low temperature is completely dominated by the 
linear part of the Bogoliubov spectrum.

Most approaches to the thermodynamics of the dilute Bose gas are mean-field
approximations
~\cite{gribben,sam0,nikj,sam, haug,henk2,jens1,jens3,japs,popov1,hk}. 
Common to all mean-field calculations is that they predict
a critical temperature $T_c$ for Bose condensation which is the same as
that of an ideal Bose gas~\cite{huang}. 
Moreover some of the mean-field approximations such as the 
Popov appoximation
predict a first-order phase transition. This is in disagreement with
universality-class arguments which tell us that the NR Bose gas goes 
through a second-order phase transition since the it belongs to the
universality class of the 3-dimensional $O(2)$-symmetric spin model. 
Another problem with some of
the mean-field approximation such as the Hartree-Fock-Bogoliubov (HFB)
approximation, is that the dispersion relation is gapped~\cite{gribben,jens2}. 
This is in conflict
with the Hugenholtz-Pines theorem~\cite{hugen,hohen} 
which states that the spontaneous breaking of the
$U(1)$ symmetry in the nonrelativistic 
(NR) Bose gas gives rise to a gapless mode.
Note however, that this problem with the HFB approximation
has recently been solved
by Yukalov and Kleinert by introducing a separate chemical potential for the
condensate~\cite{hk} (see also Ref.~\cite{iva} for a modfied gapless 
Hartree-Fock approach in the context of relativistic 
$O(N)$-symmetric theories).

Examples of calculations beyond mean-field theory are the renormalization
group calculations of the spin-zero Bose gas 
of Refs.~\cite{henk1,am,alber,kop}. 
Such calculations typically sum up classes of diagrams from all orders
of perturbation theory and they 
show that the
dilute Bose gas undergoes a second-order phase transitions in accordance
with expectation based on universality~\cite{henk1,am}. Other 
calcuations~\cite{berges,rammer} are based on the 
two-particle-irreducible (2PI) 
diagram effective-action approach first developed
by Luttinger and Ward~\cite{ward} and by Baym~\cite{baymnr} 
in the context of nonrelativistic fermions. Combined with the
$1/N$ expansion, the 2PI effective action is particular 
suited to deal with nonequilibrium phenomena~\cite{berges}, and has also
been used successfully to calculate critical exponenents~\cite{alford}.
Very recently, the equation of state of the dilute Bose gas was calculated
by Pilati {\it et al}~\cite{mc} using the Path Integral Monte Carlo method.

The problem of calculating the critical temperature for a dilute Bose gas
has been around for half a century (See. Refs.~\cite{jens2,yuk} for a review), 
butthe issue been was first settled 6-7 years ago
by Baym {\it et al}~\cite{baym1,baym2}. 
The Hartree-Fock (HF) approximation which amounts to perturbation theory to
leading order in the scattering length merely implies a redefinition of
the chemical potential and hence predicts no shift in the critical 
temperature. 
Going beyond the HF approximation using 
perturbation theory, one is facing infrared divergences 
which become more severe as one goes to higher orders.
Thus perturbation theory breaks down close to $T_c$ which shows
that the physics of the phase transition is inherently nonperturbative.
Using effective-field-theory
methods, it was shown rigorously by Baym {\it et al}~\cite{baym1,baym2}
that the parametric dependence of the change in the critical temperature,
$\Delta T_c$, is linear in the scattering length $a$
in the dilute limit:
\bqa
{\Delta T_c\over T_c^0}&=&c\left({\rho_c^0}\right)^{1/3}a\;,
\eqa
where $c$ is a constant, and $T_c^0$ and $\rho_c^0$ are 
the critical temperature and critical density of the ideal
Bose gas, respectively.
The coefficient $c$ has been determined by analytical as well as numerical
means in recent years. This includes 
effective-field theory~\cite{baym1,baym2,peter3},
high-precision Monte-Carlo
calculations~\cite{peter1,peter2,kutrus,landy}, 
simulations of classical field theory in the microcanonical 
ensemble~\cite{davmor},
$1/N$-expansions~\cite{baym3,tomasik}, 
variational perturbation theory~\cite{boris}, and the linear delta 
expansion~\cite{susa1,susa2,radu}.

The fact that mean-field theories and perturbation theory fail in the critical
region, warrants a nonperturbative approach.
In the present paper, we apply the $1/N$-expansion~\cite{brezin,zinn} 
to the interacting 
NR Bose gas at finite temperature. 
In contrast with e.g. the HFB approximation, this approach satisfies
the Hugenholtz-Pines theorem order by order in $1/N$.
This expansion 
is a nonperturbative method that sums up infrared-divergent diagrams from 
all orders of perturbation theory. 
For example, a leading-order calculation sums up all daisy and superdaisy
diagrams. Moreover,
in order to get results beyond mean field, one must typically go 
to nex-to-leading order (NLO) in the $1/N$ expansion.
Following the 
approach in Ref.~\cite{abw}, we derive the thermodynamic
quantities at finite temperature to next-to-leading order in the 
$1/N$-expansion.

The paper is organized as follows. In Sec.~II, we briefly discuss the
nonrelativistic Bose gas and the $1/N$-expansion. In Sec.~III, 
we discuss the thermodynamic potential and the gap equations. We also
derive the leading-order and 
next-to-leading-order results. We summarize in Sec.~IV.

\section{Dilute Bose Gas and Effective action}
The Euclidean Lagrangian for a nonrelativistic Bose gas with $N$ species
is
\bqa\nonumber
{\cal L}&=&\hbar\psi^{\dagger}_i\partial_{\tau}\psi_i
+{\hbar^2\over2M}\nabla\psi^{\dagger}_i
\cdot\nabla\psi_i-\mu\psi^{\dagger}_i\psi_i
+{g\over2N}(\psi^{\dagger}_i\psi_i)^2\;,
\\ &&
\label{lag}
\eqa
where $i=1,2,...,N$, $\psi_i$ is a complex field, $\mu$ is the chemical
potential\footnote{In principle, one could introduce $N$ different chemical
potentials $\mu_i$, but this is not natural in this context.
The symmetry group would then be $[O(2)]^N$.} and $M$ is the mass of the
bosons. 
In the following we will be using units where 
$2M=\hbar=1$ and summation over repeated indices
is understood. 
The Lagrangian~(\ref{lag}) is invariant under the group $O(2N)$.
For $N=1$, this reduces to the Lagrangian for the nonrelativistic Bose gas, and
we can identify the coupling constant with the $s$-wave scattering length, 
$g=8\pi a$~\footnote{Note that the expansion parameter formally is $1/2N$
so the expansion makes sense also for $N=1$.}. 
In order to eliminate the quartic interaction from the Lagrangian
we introduce an auxiliary field $\alpha$ and add to Eq.~(\ref{lag})
the following term
\bqa
{\cal L}_{\alpha}&=&{2N\over g}\left[
\alpha-{ig\over2N}\left(\psi_i^{\dagger}\psi_i-{N\mu\over g}\right)
\right]^2\;.
\eqa
The Lagrangian can now be written as
\bqa\nonumber
{\cal L}&=&{2N\over g}\alpha^2-2i\alpha\left(
\psi_i^{\dagger}\psi_{i}-{N\mu\over g}
\right)+\psi^{\dagger}_i\partial_{\tau}\psi_i
+\nabla\psi^{\dagger}_i\cdot\nabla\psi_i
\\&&
-{N\mu^2\over2g}
\;.
\label{newlag}
\eqa
By using the equation of motion for $\alpha$, one can eliminate this field
altogether and one recovers the original Lagrangian~(\ref{lag}).

Integrating over the fields $\psi_2,\psi_3...\psi_N$, we obtain the 
effective action for $\alpha$ and $\psi_1$:
\bqa\nonumber
S_{\rm eff}&=&
(N-1){\rm Tr}\ln
\left[\partial_{\tau}-\nabla^2-2i\alpha
\right]
\\ &&\nonumber
+\int_0^{\beta}d\tau\int d^3x\left[
\psi_1^{\dagger}\partial_{\tau}\psi_1
+\nabla\psi^{\dagger}_1\cdot\nabla\psi_1
+{2N\over g}\alpha^2
\right.\\&&
\left.
-2i\alpha\left(\psi_1^{\dagger}\psi_1-{N\mu\over g}
\right)
-{N\mu^2\over2g}\right]\;,
\label{ea}
\eqa
where $\beta=1/T$.
In Eq.~(\ref{ea}), ${\rm Tr}$ implies taking the trace of the
differential operator inside the brackets.
We next parametrize the fields $\psi_1$ and $\alpha$ by writing them as
sums of space-time independent expectation values $\phi_0$ and 
$im$ and quantum fluctuating fields $\tilde{\psi}$ and $\tilde{\alpha}$:
\bqa
\label{para1}
\psi_1&=&\sqrt{N}\phi_0+\tilde{\psi}_1\;,\\
\alpha&=&im+{\tilde{\alpha}\over\sqrt{N}}\;.
\label{para2}
\eqa
The rescaling with factors of $1/\sqrt{N}$ is just facilitate the 
counting of factors of $N$.
The field $\phi_0$ can be rotated so it is real.
It can be shown that the expectation value of $\alpha$ is purely imaginary
so $m$ is real.
Substituting Eqs.~(\ref{para1})--(\ref{para2}) into Eq.~(\ref{ea}), we obtain
\bqa\nonumber
S_{\rm eff}&=&
(N-1){\rm Tr}\ln
\left[\partial_{\tau}-\nabla^2+2m
-{2i\tilde{\alpha}\over\sqrt{N}}
\right]
\\ &&\nonumber
+\int_0^{\beta}d\tau\int d^3x\left[
\tilde{\psi}_1^{\dagger}\partial_{\tau}\tilde{\psi_1}
-\nabla\tilde{\psi}_1^{\dagger}\cdot\nabla\tilde{\psi}_1
\right.
\\&& \nonumber
+{2N\over g}\left(im+{\tilde{\alpha}\over\sqrt{N}}\right)^2
+2\left(m-{i\tilde{\alpha}\over\sqrt{N}}
\right)
\\ &&\left.
\nonumber
\times\left(N\phi_0^2+\sqrt{N}\phi_0(\tilde{\psi}_1^{\dagger}+\tilde{\psi}_1)
+\tilde{\psi}_1^{\dagger}\tilde{\psi}_1
\right.\right.
\\ &&\left.\left.
-{N\mu\over g}\right)
-{N\mu^2\over2g}
\right]\;,
\label{ea2}
\eqa
where we have set $k_B=1$.
Writing $\tilde{\psi}_1=(\phi_1+i\phi_2)/\sqrt{2}$ and
expanding Eq.~(\ref{ea2}) in powers of $1/\sqrt{N}$ through order
$1/N$, we obtain
\bqa\nonumber
{S_{\rm eff}\over\beta V}&=&
(N-1)\sumint_P\ln\left[
ip_0+p^2+2m
\right]
+2mN\phi_0^2
\\&&\nonumber
-N{(\mu+2m)^2\over2g}
\\ &&\nonumber
\hspace{-1.4cm}
+{1\over2}\sumint_P\chi^T(-P)
\left(\begin{array}{ccc}
p^2+2m&p_0&-2\sqrt{2}i\phi_0\\
-p_0&p^2+2m&0\\
-2\sqrt{2}i\phi_0&0&{4\over g}+4\Pi(P,m)
\end{array}\right)\chi(P)\;, 
\\ 
&&
\label{1n}
\eqa
where $V$ is the volume of the system, 
$\chi=(\phi_1(P),\phi_2(P),\tilde{\alpha}(P))$ 
are the Fourier transforms of $\phi_1$, $\phi_2$, and $\tilde{\alpha}$.
We have neglected linear terms that vanish at the minimum of the
thermodynamic potential. 
We have here introduced the sum-integral
\bqa
\sumint_Q&\equiv&T\sum_{q_0=2\pi nT}\int{d^3q\over(2\pi)^3}\;.
\eqa
The integral over three-momentum is regularized by using a three-dimensional
cutoff $\Lambda$.
Finally, the function $\Pi(P,m)$ is defined by
\bqa\nonumber
\Pi(P,m)&=&
\sumint_Q{1\over q^2+2m+iq_0}{1\over (p+q)^2+2m-i(p_0+q_0)}\;.
\\ &&
\eqa

\section{Thermodynamic potental and Gap equations}
The thermodynamic potential ${\Omega}$ 
is given by all one-particle irreducible diagrams and 
is given by a series in $1/N$. The
first two
terms are easily obtained from Eq.~(\ref{1n}) by gaussian integration over
$\chi$. We can then write
\bqa
\Omega&=&N\Omega_{\rm LO}+\Omega_{\rm NLO}
\;,\eqa
where
\bqa\nonumber
\Omega_{\rm LO}&=&
-{\left(\mu+2m\right)^2\over2g}
+2m\phi_0^2
\\ &&
+
\sumint_P\ln\left[ip_0+p^2+2m\right]\;,
\label{loom}
\\ \nonumber
\Omega_{\rm NLO}&=&
{1\over2}\sumint_P\ln\left[
\Pi(p,m)+{1\over g}+{2\phi_0^2(p^2+2m)\over p_0^2+(p^2+2m)^2}
\right]\;.
\\ &&
\eqa
The free energy ${\cal F}$ is given by all connected diagrams and is 
independent of the condensate $\phi_0$ and the expectation value $m$.
So if the thermodynamic potential is evaluated at the values of 
$\phi_0$ and $m$, which satisfy
the equations
\bqa
\label{gap1}
{\partial \Omega\over\partial\phi_0}&=&0\;,\\
{\partial \Omega\over\partial m}&=&0\;,
\label{gap2}
\eqa
the one-particle reducible diagrams vanish.
The values of $\phi_0$ and $m$, which satisfy the stationarity 
conditions~(\ref{gap1}) and~(\ref{gap2}) can be written as series in $1/N$:
\bqa
\phi_0&=&\phi_{0,\;\rm LO}+{1\over N}\phi_{0,\;{\rm NLO}}
\;, \\
m&=&m_{\rm LO}+{1\over N}m_{\rm NLO}\;.
\eqa
By series expanding the thermodynamic potential around the LO solution to the
gap equations, the free energy can then be written as
\bqa\nonumber
{\cal F}&=&N\Omega_{\rm LO}(m_{\rm LO},\phi_{0,\;{\rm LO}})+
\Omega_{\rm NLO}(m_{\rm LO},\phi_{0,\;{\rm LO}})
\\ && \nonumber
+m_{\rm NLO}
{\partial \Omega_{\rm LO}\over\partial m}
\bigg|_{m=m_{\rm LO},\phi_0=\phi_{0,\;{\rm LO}}}
\\ &&
+\phi_{0,\;\rm NLO}
{\partial \Omega_{\rm LO}\over\partial\phi_0}
\bigg|_{m=m_{\rm LO},\phi_0=\phi_{0,\;{\rm LO}}}
+{\cal O}\left(1/N^2\right)\;.
\eqa
The free energy to NLO then reduces to 
\bqa
{\cal F}&=&N{\cal F}_{\rm LO}+{\cal F}_{\rm NLO}\;,
\eqa
where
\bqa
{\cal F}_{\rm LO}&=&\Omega_{\rm LO}(m_{\rm LO},\phi_{0,\;{\rm LO}})
\;,\\
{\cal F}_{\rm NLO}&=&\Omega_{\rm NLO}(m_{\rm LO},\phi_{0,\;{\rm LO}})\;.
\eqa
The number density $\rho$ is given by the expectation value
\bqa
\rho&=&\langle\psi^*\psi\rangle\;.
\eqa
From the path-integral representation of the free energy $\cal F$
\bqa
e^{-\beta V\cal F}&=&\int{\cal D}\psi^*{\cal D}\psi 
e^{-\int_0^{\beta}d\tau\int d^3x\cal L}\;,
\eqa
and Eq.~(\ref{lag}), the expression for the number density can be written as
\bqa
\rho&=&-{\partial \cal F\over\partial\mu}\;.
\label{dens}
\eqa
Using Eqs.~(\ref{newlag}) and~(\ref{dens}), one finds the following exact
result
\bqa
\rho&=&N{\mu+2m\over g}\;.
\label{exact}
\eqa

\subsection{Leading-order results}
At leading order the gap equations~(\ref{gap1}) and~(\ref{gap2}) become
\bqa
\label{logap1}
4m\phi_0&=&0\;,\\
\phi_0^2&=&{\mu+2m\over g}-\sumint_P{1\over ip_0+p^2+2m}\;.
\label{logap2}
\eqa
The solution to Eq.~(\ref{logap1}) is that either $m$ or $\phi_0$ vanishes.
$m=0$ corresponds to the Bose-condensed phase and we discuss this first.

Substituting $m=0$ into Eq.~(\ref{loom}),
the leading-order contibution to the free energy reduces to
\bqa\nonumber
{\cal F}_{\rm LO}&=&
-{\mu^2\over2g}
+\sumint_P\ln\left[ip_0+p^2\right]\;.
\eqa
The sum-integral can 
be calculated by first writing the sum over Mastubara frequencies as 
a contour integral. By performing this contour integral and integrating over
angles, we are left with
\bqa\nonumber
{\cal F}_{\rm LO}&=&
-{\mu^2\over2g}+
{1\over4\pi^2}\int_0^{\Lambda}dp\;p^2\left[p^2+2T\ln\left(
1-e^{-\beta p^2}\right)\right]\\
&=&-{\mu^2\over2g}
+{\Lambda^{5}\over20\pi^2}
-{4\pi}\left({T\over4\pi}\right)^{5/2}\zeta\left(\mbox{$5\over2$}\right)
\;.
\label{lofr}
\eqa
The divergent term is eliminated by adding a 
vacuum counterterm $\Delta{\cal F}$ to the free energy $\cal F$ and
the renormalized LO free energy is
\bqa
{\cal F}_{\rm LO}
&=&-{\mu^2\over2g}
-{4\pi}\left({T\over4\pi}\right)^{5/2}\zeta\left(\mbox{$5\over2$}\right)
\;.
\label{rlofr}
\eqa
Note that the LO free energy density does not agree with the standard
mean-field result. The first term in Eq.~(\ref{rlofr}) is the zero-temperature
result in the mean-field approximation, while the second term is the
finite-temperature contribution for an ideal Bose gas.
The LO number density in terms of the condensate $\phi_0$ 
follows from Eqs.~(\ref{exact}) and~(\ref{logap2})
\bqa
\rho&=&\phi_0^2+\sumint_P{1\over ip_0+p^2}
\;.
\label{div}
\eqa
The sum-integral in the gap equation~(\ref{div}) can be calculated 
in a similar manner as the one appearing in the free energy and one obtains
\bqa
\rho&=&\phi_0^2
+{\Lambda^3\over12\pi^2}
-\left({T\over4\pi}\right)^{3/2}\zeta\left(\mbox{$3\over2$}\right)
\;.
\label{fphi}
\eqa
The divergent term reflects the fact that the expectation value of the
number density operator also has divergences and we need to add a counterterm
$\delta\rho$ to eliminate them~\cite{eric}. This yields
\bqa
\label{rhoc}
\rho&=&\phi_0^2+\left({T\over4\pi}\right)^{3/2}\zeta
\left(\mbox{$3\over2$}\right)
\;.
\eqa
The critical temperature is found by setting $\phi_0=0$.
This yields
\bqa
T_c^0&=&4\pi\left[{n\over\zeta\left(\mbox{$3\over2$}\right)}\right]^{2/3}
\;,
\label{rhoc2}
\eqa
which is the standard result for an ideal Bose gas~\cite{huang}.
Inverting
Eq.~(\ref{rhoc2}) then gives the critical number density $\rho_c^0$ 
for Bose condensation
\bqa
\rho_c^0&=&\zeta\left(\mbox{$3\over2$}\right)
\left[{T\over4\pi}\right]^{3/2}
\;.
\label{idealrhoc}
\eqa

We next discuss the symmetric phase where $\phi_0=0$.
The free energy in the symmetric phase is given by
\bqa\nonumber
{\cal F}&=&
\sumint_P\ln\left[ip_0+p^2+2m\right]
-{1\over2}g\left(\sumint_P{1\over ip_0+p^2+2m}\right)^2\\ 
&=&\nonumber
-{4\pi}\left({T\over4\pi}\right)^{5/2}
{\rm Li}_{5/2}\left(e^{-2\beta m}\right)
\\ &&
-{1\over2}g\left({T\over4\pi}\right)^{3}
{\rm Li}_{3/2}^2\left(e^{-2\beta m}\right) 
\label{freee}
\;,
\eqa
where ${\rm Li}_{n}(x)=\sum_kx^k/k^n$ is the polylogarithmic function
and we
have have eliminated $\mu+2m$ in Eq.~(\ref{loom}) using Eq.~(\ref{gap2}).
In the symmetric phase, the number density reduces to
\bqa\nonumber
\rho&=&
\sumint_P{1\over ip_0+p^2+2m}
\\ 
&=&\left({T\over4\pi}\right)^{3/2}{\rm Li}_{3/2}\left(e^{-2\beta m}\right)
\;,
\eqa
where we have neglected a temperature-independent divergent term.
The number density in the symmetric phase can be interpreted as
that of an ideal Bose gas with an effective chemical potential 
$\mu_{\rm eff}=-2m$. Eq.~(\ref{logap2}) can be interpreted as
a self-consistent one-loop equation
for the chemical potential and the critical number
density for Bose condensation is then given by $\mu_{\rm eff}=0$.

\subsection{Next-to-leading order results}
The NLO gap equations~(\ref{gap1})--(\ref{gap2}) are
\bqa\nonumber
0&=&\left[m+{1\over2N}\sumint_P
{1\over\Pi(P,0)+{1\over g}+{2\phi_0^2p^2\over p_0^2+p^4}}{p^2\over p_0^2+p^4}
\right]\phi_0\;, 
\\&&
\label{1om}
\\\nonumber
\phi_0^2&=&{\mu+2m\over g}-\sumint_P{1\over ip_0+p^2+2m}
\\&& \nonumber
-{1\over N}\sumint_P
{1\over\Pi(P,0)+{1\over g}+{2\phi_0^2p^2\over p_0^2+p^4}}
\\&& 
\times\left[{1\over4}{d\Pi(P,0)\over dm}
+{\phi_0^2\over p_0^2+p^4}
-{2p^4\phi_0^2\over(p_0^2+p^4)^2}
\right]\;,
\label {logap3}
\eqa
where we have used that it is consistent to set $m=0$ in the NLO terms.
Setting $N=1$, 
the expression for the free energy now becomes
\bqa\nonumber
{\cal F}&=&
-{\mu^2\over2g}
+
\sumint_P\ln\left[ip_0+p^2\right]
\\ &&+
{1\over2}
\sumint_P\ln\left[
\Pi(p,0)+{1\over g}+{2\phi_0^2p^2\over p_0^2+p^4}
\right]\;,
\label{fnlo}
\eqa
where $\phi_0$ satisfies the LO gap equation~(\ref{logap2}) and is to be
considered a function of $\mu$. The number density can be derived from~(\ref{fnlo})
and reads
\bqa
\rho&=&{\mu\over g}+{1\over g}\sumint_P
{1\over\Pi(P,0)+{1\over g}+{2\phi_0^2p^2\over p_0^2+p^4}}{p^2\over p_0^2+p^4}
\eqa
Another more useful expression for the number density follows from 
Eqs.~(\ref{exact}) and~(\ref{logap3})
\bqa\nonumber
\rho&=&
\phi_0^2
+\sumint_P{1\over ip_0+p^2+2m}
\\&& \nonumber
+\sumint_P
{1\over\Pi(P,0)+{1\over g}+{2\phi_0^2p^2\over p_0^2+p^4}}
\\&& 
\times\left[{1\over4}{d\Pi(P,0)\over dm}
+{\phi_0^2\over p_0^2+p^4}
-{2p^4\phi_0^2\over(p_0^2+p^4)^2}
\right]\;.
\label{complete}
\eqa
The $1/N$ expansion is not limited to a weakly interacting Bose gas.
If we make the additional assumption of weak coupling, we can make further
approximations by simply neglecting $g\Pi(P,m)$. 
This approximation is valid outside the critical
region and the free energy reduces to
\bqa
{\cal F}&=&-{\mu^2\over2g}+{1\over2}\sumint_P\ln\left[
p_0^2+\epsilon^2(p)
\right]\;,
\label{free1}
\eqa
where
\bqa
\epsilon(p)&=&p\sqrt{p^2+2g\phi_0^2}\;,
\eqa
is the well-known Bogoliubov spectrum~\cite{bogo1}.
Similarly, the number density reduces to 
\bqa\nonumber
\rho&=&
\phi_0^2
+\sumint_P{1\over ip_0+p^2}
+g\phi_0^2\sumint_P{1\over p_0^2+\epsilon^2(p)}
{p_0^2-p^4\over p_0^2+p^4}\;.
\\ &&
\label{rhoc2}
\eqa
After performing the sum over Matsubara frequencies, we obtain
\bqa
\rho&=&\phi_0^2+{1\over4}\int{d^3p\over(2\pi)^3}
{\epsilon^2(p)/p^2+p^2\over\epsilon(p)}
\Big[1+2n(\epsilon(p))\Big]\;,
\label{rhobog}
\eqa
where $n(x)=1/(e^{\beta x}-1)$ is the Bose-Einstein distribution function.
At $T=0$, this reduces to the Bogoliubov approximation and gives the 
standard weak-coupling result for the depletion of the 
condensate~\cite{leeyang1}:
\bqa
\rho&=&\phi_0^2\left[1+{8\over3}\sqrt{\phi_0^2a^3\over\pi}\right]\;.
\eqa
At finite temperature, this approximation goes beyond the Bogoliubov 
approximation since we via Eq.~(\ref{rhoc2}) take into account the 
temperature dependence of the condensate $\phi_0$ when 
evaluating~(\ref{free1})~\cite{jens2}. 

In order to find the critical number density $\rho_c$, 
we must evaluate the right-hand side
of Eq.~(\ref{rhobog}) in the limit $\phi_0\rightarrow0$.
The Bogoliubov spectrum then reduces to the free-particle spectrum 
and we recover the mean-field result~(\ref{idealrhoc}).

\begin{figure}[htb]
\vspace{0.6cm}
\scalebox{0.4}{\includegraphics{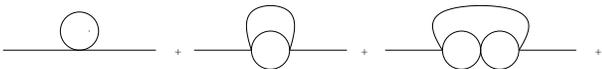}}
\caption{Diagrams contributing to the self-energy at LO in the 
$1/N$ expansion.}
\label{lself}
\end{figure}

\begin{figure}[htb]
\vspace{0.6cm}
\scalebox{0.4}{\includegraphics{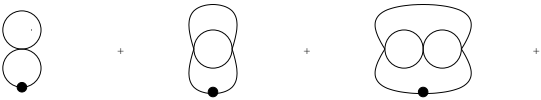}}
\caption{Diagrams contributing to the number
density at NLO in the $1/N$ expansion.}
\label{nden}
\end{figure}

We have seen that by neglecting $\Pi(P,m)$, one recovers the mean-field
results for the dilute Bose gas.
In order to go beyond mean field, we must keep the self-energy in the gap 
equations. Since the phase transition is dominated by 
the nonperturbative long-distance physics, the
dominant contribution comes from the static Matsubara 
mode~\cite{baym1,baym2,tomasik}.
The contributions from the nonstatic Matsubara modes can be treated 
perturbatively and this fact underlies the 
calculations of 
Refs.~\cite{baym1,baym2,tomasik} using dimensional-reduction techniques.
If we denote the contribution to the self-energy $\Pi(P,m)$ 
from the zeroth Matsubara frequency at $p_0=0$ by  
$\Pi_0(p,m)$, we have
\bqa
\Pi_0(p,0)&=&T\int{d^3q\over(2\pi)^3}
{1\over q^2}{1\over({\bf p}+{\bf q})^2}\;,\\
{d\Pi_0(p,m)\over dm}\bigg|_{m=0}
&=&-8T\int{d^3q\over(2\pi)^3}{1\over q^4}{1\over({\bf p}+{\bf q})^2}\;.
\label{2nd}
\eqa
The critical number density is then found by settting $\phi_0=0$ and 
inserting~(\ref{2nd}) into~(\ref{complete}) and rearranging terms. This yields
\bqa\nonumber
\rho_c&=&
\sumint_P{1\over ip_0+p^2}
-2T^2
\int{d^3q\over(2\pi)^3}{1\over q^4}
\\ &&
\times
\int{d^3p\over(2\pi)^3}{1\over\Pi_0(p,m)+1/g}{1\over({\bf p}+{\bf q})^2}\;.
\label{rhoc3d}
\eqa
The term 
\bqa
\Sigma(q)&=&
\int{d^3p\over(2\pi)^3}
{1\over\Pi_0(p,m)+1/g}{1\over({\bf p}+{\bf q})^2}
\eqa
can be interpreted as the leading correction to the self-energy function 
in the $1/N$ expansion. The corresponding string of Feynman diagrams is 
shown in Fig.~\ref{lself}. 
The first of these diagrams corresponds to the 
HF approximation. The self energy needs mass renormalization, which is
carried out by subtracting $\Sigma(0)$~\cite{zinn1}. 
Similarly, 
the second term in Eq.~(\ref{rhoc3d}) can be interpreted in terms of Feynman
diagrams~\cite{baym3,tomasik}. These are shown in Fig.~\ref{nden}.

The first term in ~(\ref{rhoc3d}) is $\rho_c^0$, while the
the second term was first calculated by 
Baym, Blaizot, and Zinn-Justin~\cite{baym3}, and later
by Arnold and Tom\'a\'sik~\cite{tomasik}. 
It is given by $gT/32\pi^2$ and the critical number density reduces to
\bqa\nonumber
\rho_c&=&\rho_c^0-{1\over2}g\left({T\over4\pi}\right)^2 \\
&=&\rho_c^0\left[
1-{4\pi\left(\rho_c^0\right)^{1/3}g\over\left[\zeta({\rm3\over2})\right]^{4/3}}
\right]
\label{finale}
\;,
\eqa
where we have used $g=8\pi a$.
Thus the critical number density decreases linearly with the scattering length
$a$. 
Instead of calculating the critical number density at fixed
temperature, we can calculate the critical temperature at fixed number density
by using 
\bqa
{\Delta T_c\over T_c^0}&\simeq&-{2\over3}{\rho_c-
\rho_c^0\over\rho_c^0}.
\;,
\eqa
which is valid to leading order in the 
number-density expansion~\cite{baym2,tomasik}.
This yields
\bqa\nonumber
{\Delta T_c\over T_c^0}&=& 
{8\pi\over3\left[\zeta(\rm{3\over2})\right]^{4/3}}\left(\rho_c^0\right)^{1/3}a
\\
&\approx&2.33
\left(\rho_c^0\right)^{1/3}
a\;. 
\label{finale2}
\eqa
It is interesting to note~\cite{baym3} that the results~(\ref{finale}) 
and~(\ref{finale2}) are
independent of $N$, although they are correct only in the limit 
$N\rightarrow\infty$.
The value 2.33 of the constant $c$ compares reasonably well with 
$c\approx1.3$ from lattice calculations~\cite{peter1,peter2,kutrus,landy}
and classical field-theory simulations~\cite{davmor}.

\section{Summary}
In this paper, we have used the $1/N$-expansion to derive thermodynamic
quantities
of the interacting Bose gas. If we make the additional assumption of weak
coupling, we obtain the standard results for the dilute Bose gas.
If we neglect the contribution to the number density from the nonstatic
Matsubara modes at criticality, we obtain the result for the critical
number density first obtained by Baym {\it et al}~\cite{baym1,baym2}
The $1/N$ expansion gives a coherent description of the dilute Bose gas
in the critical region as well as outside it, where mean-field theories
are normally sufficient.

We have not calculated the critical exponents associated with the
second-order phase transition of the Bose gas. However, making the same
approximations as we did for our calculations of the critical number
density,
one expects to obtain the standard results for the 1PI
effective action to NLO in $1/N$~\cite{zinn1}.

It is natural to ask whether the $1/N$ expansion is reliable for $N=1$.
In the case of the critical temperature for Bose condensation, the 
$1/N$ correction was calculated by Arnold and Tomasik~\cite{tomasik} and
the correction was approximately 26\%. This is as expected:
a correction of $1/2N$ or 50\% multiplied by a constant of order one.

\section*{Acknowledgment}
The author would like to thank H. J. Warringa for useful discussions and
suggestions.

\appendix
\renewcommand{\theequation}{\thesection.\arabic{equation}}



\begin{thebibliography}{99}
\bibitem{bogo1}N. N. Bogoliubov, J. Phys (Moscow) {\bf 11}, 23 (1947).
\bibitem{leeyang1}T. D. Lee and C. N. Yang, Phys. Rev. {\bf 105}, 1119 (1957).
\bibitem{leeyang2}T. D. Lee and C. N. Yang, Phys. Rev. {\bf 105}, 1419 (1958).
\bibitem{popov1}V. N. Popov, {\it functional Integrals in
Quantum Field Theory and Statistical Physics}, (Reidel,Dordrecht) (1983).

\bibitem{gribben}A. Griffin, Phys. Rev. B {\bf 53}, 9341 (1996).

\bibitem{henk2}M. Bijlsma, and H. T. C. Stoof, 1996, Phys. Rev. {\bf A55}, 
498 (1996).
\bibitem{haug}T. Haugset, H. Haugerud, and F. Ravndal, Ann. Phys. (N.Y.)
{\bf 34}, 321 (1998). 
\bibitem{nikj} N. P. Proukakis, S. A. Morgan, S. Choi, and K. Burnett
Phys. Rev. A {\bf 58}, 2435 (1998). 

\bibitem{sam0}D. A. W. Hutchinson, K. Burnett, R. J. Dodd, S. A. Morgan, 
M. Rusch, E. Zaremba, N. P. Proukakis, M. Edwards, and C. W, Clark,
J. Phys. B {\bf 33}, 3825 (2000).
\bibitem{sam}S. A. Morgan, J. Phys. B {\bf 33}, 3847 (2000).

\bibitem{jens1}J. O. Andersen, U. Al Khawaja, and H. T. C. Stoof,
Phys. Rev. Lett. {\bf 88}. 070407 (2002).
\bibitem{jens3} U. Al Khawaja, J. O. Andersen, N. P. Proukakis, 
and H. T. C. Stoof, Phys. Rev. {\bf A66}, 013615 (2002).


\bibitem{japs}T. Kita, J. Phys. Soc. Jpn. {\bf 75}, 044603 (2006).

\bibitem{hk}V. I. Yukalov and H. Kleinert, Phys. Rev. {\bf A73}, 063612 (2005).
\bibitem{huang}K. Huang, {\it Statistical Mechanics}, John Wiley \& Sons (1987).
\bibitem{jens2}J. O. Andersen,  Rev. Mod. Phys. {\bf 76}, 599 (2004).

\bibitem{hugen}N. M. Hugenholtz, and D. Pines, Phys. Rev. {\bf 116}, 489 
(1959).
\bibitem{hohen}P. H. Hohenberg, and P. H. Martin, Ann. Phys. (N.Y.) {\bf 34}, 
291 (1965).




\bibitem{iva}Yu. B. Ivanov, F. Riek, and J. Knoll, Phys.Rev. D {\bf 71},105016 (2005).
\bibitem{henk1}M. Bijlsma, and H. T. C. Stoof, 1996, Phys. Rev. {\bf A54}, 
5085 (1996). 
\bibitem{am}J. O . Andersen and M. Strickland, Phys. Rev. {\bf A60}, 1442 
(1999).
\bibitem{alber} G. Metikas and and G. Alber, Phys. B, {\bf 35} 4223 (2002).
\bibitem{kop}N. Hasselmann, S. Ledowski, and P. Kopietz, Phys. Rev. {\bf A70}, 
063621 (2004).

\bibitem{rammer}E. Lundh and J. Rammer, Phys. Rev. A {\bf 66} 033607 (2002).

\bibitem{berges}T. Gasenzer, J. Berges, M. G. Schmidt, and M. Seco, Phys. 
Rev. {\bf A72}, 063604 (2005).
\bibitem{ward}J. M. Luttinger and J. D. Ward, Phys. Rev. {\bf 118}, 1417 
(1960).
\bibitem{baymnr}G. Baym, Phys. Rev. {\bf 127}, 1391 (1962). 
\bibitem{alford}M. Alford, J. Berges, and J. M. Cheyne, 
Phys. Rev. D {\bf 70}, 125002 (2004).
\bibitem{mc}S. Pilati, K. Sakkos, J. Boronat, J. Casulleras, and S. Giorgini,
cond-mat/0607721.
\bibitem{yuk}V. I Yukalov, Laser Phys. Lett. {\bf 1}, 435 (2004).
 
\bibitem{baym1}
G. Baym, J.-P. Blaizot, M. Holzmann, F. Lal\"oe, and D. Vautherin,
Phys. Rev. Lett. {\bf 83}, 1703 (1999).
\bibitem{baym2}
G. Baym, J.-P. Blaizot, M. Holzmann, F. Lal\"oe, and D. Vautherin,
Eur. Phys. J. B {\bf 24}, 107 (2001). 
\bibitem{peter3}P. Arnold, G. M. Moore, and B. Tomasik, Phys. Rev. A{\bf 65}, 0130606 (2002).
\bibitem{peter1}P. Arnold and G. M. Moore, Phys. Rev. Lett. {\bf 87}, 120401
(2001).
\bibitem{peter2}P. Arnold and G. M. Moore, Phys. Rev. E {\bf 64} 066113 (2001).

\bibitem{kutrus}V. A. Kashurnikov, N. V. Prof'ev, and B. V. Svistunov,
Phys. Rev. Lett. {\bf 87}, 120402.
\bibitem{landy}K. Nho and D. P. Landau, Phys. Rev. Lett. {\bf 90}, 040402 
(2003).
\bibitem{davmor}M. J. Davis and S. A. Morgan, Phys. Rev. A {\bf 68}, 
053615 (2003).
\bibitem{baym3}G. Baym, J.-P. Blaizot, and J. Zinn-Justin, Europhys. Lett. {\bf 49}, 150 (2000).
\bibitem{tomasik}P. Arnold and B. Tomasik, Phys. Rev. A
{\bf 62}, 063604 (2000).


\bibitem{boris}B. Kastening, Phys. Rev. A {\bf 69}, 043613 (2004).


\bibitem{radu}E. Braaten and E. Radescu, Phys. Rev. A {\bf 66}. 063601 (2002).
\bibitem{susa1}F. F. de Souza Cruz, F. M. B. Pinto, and P. O. Ramos, Phys. 
Rev. B {\bf 64}, 014515 (2001).
\bibitem{susa2}F. F. de Souza Cruz, F. M. B. Pinto, P. O. Ramos, and P. Sena 
Phys. Rev. A {\bf 65}, 053613 (2002).



\bibitem{brezin}E. Bre\`zin and S. R. Wadia, Eds. {\it The Large-N
Expansion in Quantum Field Theory and Statistical Physics} (World Scientific,
Singapore) (1993).
\bibitem{zinn}M. Moshe and J. Zinn-Justin Phys. Rep. {\bf 385} 69 (1998).
\bibitem{abw}J. O. Andersen, D. Boer, and H. J. Warringa, Phys. Rev. 
D {\bf 69}, 076006 (2004); ibid D {\bf 70}, 116007 (2004).
\bibitem{eric}E. Braaten and A. Nieto, Eur. Phys. J. B {\bf 11}, 143 (1999).
\bibitem{zinn1}J. Zinn-Justin, {\it Quantum Field Theory and Critical 
Phenomena} (Oxford University, New York) (1989).










\end{thebibliography}
\end{document}